# Photon conservation in trans-luminal metamaterials

J. B. Pendry[1] *, E. Galiffi[1,2], and P. A. Huidobro[3]

[1]*The Blackett Laboratory, Department of Physics, Imperial College London, London, SW7 2AZ, UK .*
[2]*Advanced Science Research Center, City University of New York, 85 St. Nicholas Terrace, 10031 New York, NY, USA.*
[3]*Instituto de Telecomunicações, Instituto Superior Tecnico-University of Lisbon, Avenida Rovisco Pais 1, Lisboa, 1049-001 Portugal.*
*\*j.pendry@imperial.ac.uk*

**Abstract:** Structures which appear to move at or near the velocity of light contain singular points. Energy generated by the motion accumulates at these points into ever-narrowing peaks. In this paper we show that energy is generated by a curious process that conserves the number of photons, adding energy by forcing photons already present to climb a ladder of increasing frequency. We present both a classical proof based on conservation of lines of force, and a more formal QED proof demonstrating the absence of unpaired creation and annihilation operators. Exceptions to this rule are found when negative frequencies make an appearance. Finally we make a connection to laboratory-based models of black holes and Hawking radiation.

## 1. Introduction

In previous work we identified a novel gain mechanism for light which is present in structures synthetically moving close to the velocity of light, trans-luminal structures as we shall call them. This mechanism is distinct from other gain processes such as parametric amplification, or gain media where excited atoms transfer energy to a coherent wave. These well know processes amplify light by adding more photons to the system. Here we show that, in contrast, the trans-luminal mechanism conserves the number of photons and energy is added by raising the frequency of the photons. This result is addressed both in a semi-classical context exploiting our previous result stating that lines of force are conserved, and in the context of QED where photon number conservation arises in a natural fashion for impedance-matched systems. Connections with singular gravitational metrics are made and some remarks relevant to possible models for laboratory-based black hole analogues.

## 2. Setting the scene

There is growing interest in electromagnetic properties of time-dependent structures [1], particularly where time dependence involves no physical motion of material, but rather phased modulation. Whilst pioneering works on such spatiotemporally modulated media date as early as the 1960s [2,3], the discovery of novel effects and opportunities for applications such as nonreciprocity [4-7], amplification [8-11], optical drag [12-14] and topology [15-19] has recently revived interest in the field [20]. A simple realisation might be some fixed modulation of one or both constitutive parameters of the form $\varepsilon(x - c_g t)$. There being no physical motion the velocity, $c_g$, is unrestricted in its magnitude and by varying from zero to infinity can encompass both pure spatial and purely temporal modulations. We shall be interested in the trans-luminal region where the speed of light is comparable to the grating velocity, $c_0 \approx c_g$.

In general these systems, though they break time reversal symmetry, are parity-time (PT) symmetric and in the case of a periodic structure the usual features of Bloch bands punctuated by gaps appear. For super-luminal velocities there is a twist to the story: rather than the familiar gaps in frequency, gaps appear in wave vector. In these super-luminal gaps waves gain energy as time progresses and photons are added to the system by parametric amplification. However Bloch symmetry fails in the transluminal regime [21].

Consider a wave propagating along the $x$ - direction in a periodic structure described by,

$$\varepsilon(x - c_g t) = \varepsilon_1 + 2\alpha_\varepsilon \cos(gx - \Omega t)$$
$$\mu(x - c_g t) = \mu_1 + 2\alpha_\mu \cos(gx - \Omega t) \quad (1)$$

This model has been widely adopted in other studies and shows all the rich structure described above. We use the model for illustrative purposes but our conclusions will have wider validity.

If we choose to impedance match the structure,

$$\frac{\mu(x - c_g t)}{\varepsilon(x - c_g t)} = Z^2 \quad (2)$$

where $Z$ is a constant, then all back scattering is eliminated as Maxwell's equations factorise into independent forward and backward travelling waves. This results in the closure of all band gaps and elimination of parametric amplification. Waves move forward continuously their velocity, $c_\ell$, varying according to the local value of the refractive index, $n = \sqrt{\varepsilon\mu}$. However a curious anomaly remains in the trans-luminal region defined by a grating speed that lies between the maximum and minimum local wave velocities. We illustrate this case in Fig. 1. Here we show a trans-luminal grating which is partitioned into two parts one where the local velocity of light is less than that of the grating, $c_\ell < c_g$, the other where $c_\ell > c_g$.

Consider for a moment a different case where $c_\ell \gg c_g$ or $c_\ell \ll c_g$. There being no partitions a pulse of light would move smoothly through the grating, accelerating and decelerating as it went. There is also gain and loss at work due to the time dependence. Gain regions are shown in red in the figure, loss regions in cyan. Spending equal time in each region the pulse grows and dies with no net gain, so the systems remains in a stable PT-symmetric phase. In contrast for the example we show, pulses are trapped in their respective partitions from which they cannot escape. A pulse trapped between $X_1$ and $X_2$ travels faster than the grating and moves towards $X_2$ and into the red region. It cannot pass $X_2$ which it approaches ever more slowly, compacting as it does so, and growing in amplitude because this is a gain region. On the other hand a pulse trapped between $X_2$ and $X_3$ is overtaken by the grating, moving towards $X_2$ where it compresses and gains energy.

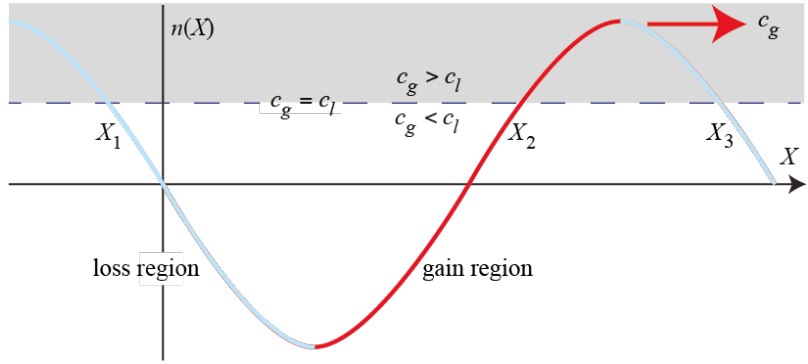

Fig. 1. Variation of refractive index, $n$, taken from Eq.(1) plotted as a function of $X = x - c_g t$. The local velocity of light, $c_\ell$, varies with $n(X)$ and in this example the grating is trans-luminal so that in one part of the grating (white-background area under the horizontal dashed line) light travels faster the grating, and in the other part (gray-shaded area) it travels slower than the grating. Gain and loss are present, gain indicated in red, loss in cyan, due to the gradient of the refractive index.

This is the novel mechanism for gain of which we spoke in earlier papers [8,10,21,22], quite distinct from parametric gain which is eliminated from our model through impedance matching. We showed in these papers that a curious conservation law holds: the number of lines of force associated with the trapped pulses is conserved even though their energy content increases exponentially with time [10,22]. The result applies generally to impedance-matched systems. For example, trapping can occur in an isolated oscillation of the refractive index, just as in a periodic system. The result is valid to a good degree of accuracy whenever back scattering can be neglected.

In the next section we go on to show that conservation of lines of force inevitably implies conservation of the number of photons. Energy can only be injected by raising the frequency of the photons.

### 3. Photon conservation

In previous papers we showed that in the absence of back scattering, time dependent structures can amplify light but without adding lines of force to existing electromagnetic fields. Here we show that conservation of lines of force necessarily implies photon conservation. First we postulate a distortion of the distribution of lines of force, which is then Fourier analysed into frequencies and finally the photon number counted by summing over frequencies.

We consider a plane wave incident on a time-dependent system such as described above, and assume no back scattering so that our conservation theorem holds exactly. The incident electric field has the form,

$$D = D_0 e^{ikx - i\omega t}, \quad \omega = c_0 k \qquad (3)$$

The number of photons in one period, $a$, is given by the energy content of the period divided by the photon energy,

$$N_{ph} = \frac{a D_0^2}{\varepsilon_0 \hbar |\omega|} = \frac{a D_0^2}{\varepsilon_0 c_0 \hbar |k|} \qquad (4)$$

Suppose that waves emerge into vacuum from the far side of the system with their fields subject to an arbitrary periodic compression defined by $f(x)$ such that,

$$f(x = na) = na \tag{5}$$

where $n$ is an integer and $a$ is the periodicity. Typically $f(x)$ might look like the red curve shown in Fig. 2.

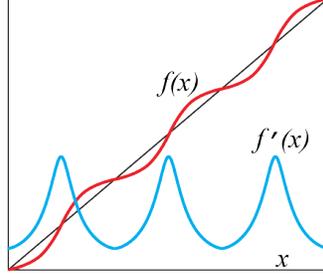

Fig. 2. A periodic compression of lines of force defined by $f'(x)$ (blue), the derivative of $f(x)$ (red), shown here over 3 periods. $f'(x)$ also defines compression of the phase.

After noting that the field lines and phase are compressed in the same fashion as the field [21], it follows that,

$$D = f'(x) D_0 e^{ikf(x)} \tag{6}$$

The model can easily be generalised to non-periodic distributions but is simplest to present in this format.

We decompose the compressed field into a set of plane waves,

$$f'(x) e^{ikf(x)} D_0 = \sum_{n=-\infty}^{+\infty} d_n e^{ikx + i2\pi nx/a} \tag{7}$$

where,

$$d_n = \frac{1}{a} \int_0^a f'(x) D_0 e^{ikf(x) - ikx - in2\pi x/a} dx = \frac{k + n2\pi/a}{ka} \int_0^a D_0 e^{ikf(x) - ikx - in2\pi x/a} dx \tag{8}$$

Next we calculate the number of photons in period $a$ remembering that negative frequencies have positive energies,

$$N_{ph} = \sum_{n=-\infty}^{+\infty} \frac{a |d_n|^2}{\hbar \varepsilon_0 c_0 |k + n2\pi/a|}$$

$$= \frac{D_0^2}{\hbar \varepsilon_0 c_0} \sum_n \frac{|k + n2\pi/a|}{k^2 a} \iint e^{ik(f(x) - f(x')) - i(k + n2\pi/a)(x - x')} dx dx' \tag{9}$$

where $\hbar |k + n2\pi/a| / \sqrt{\varepsilon_0 \mu_0}$ is the energy of a photon in the nth state.

Note that,

$$\sum_{n=-\infty}^{+\infty} |k+n2\pi/a| e^{-i(k+n2\pi/a)x} = \begin{bmatrix} +\sum_{n=n_0}^{\infty} (k+n2\pi/a) e^{-i(k+n2\pi/a)x} \\ -\sum_{n=-\infty}^{n_0-1} (k+n2\pi/a) e^{-i(k+n2\pi/a)x} \end{bmatrix}$$

$$= i\frac{d}{dx}\left[ +\sum_{n=n_0}^{\infty} e^{-i(k+n2\pi/a)x} - \sum_{n=-\infty}^{n_0-1} e^{-i(k+n2\pi/a)x} \right] \qquad (10)$$

$$= i\frac{d}{dx} e^{-i(k+n_0 2\pi/a)x} \left[ +\sum_{n=0}^{\infty} e^{-in2\pi/ax} - \sum_{n=1}^{\infty} e^{+in2\pi/ax} \right]$$

where $n_0$ is chosen so that the modulus requirement is always fulfilled. Since we are concerned with forward travelling waves, $k > 0$, it follows that $n_0 \leq 0$.

The summation can be performed,

$$+\sum_{n=0}^{\infty} e^{-in2\pi/ax} - \sum_{n=1}^{\infty} e^{+in2\pi/ax} = \left[\frac{\cos(\pi/ax)}{i\sin(\pi/ax)} + 1\right] = \frac{e^{+i\pi/ax}}{i\sin(\pi/ax)} \qquad (11)$$

and on substituting into (9) and integrating by parts,

$$N_{ph} = \frac{D_0^2}{\hbar\varepsilon_0 c_0} \frac{1}{k^2 a} \iint ikf'(y+x) e^{ik(f(x)-f(y+x))} \left[\frac{e^{iy(k+n_0 2\pi/a-\pi/a)}}{\sin(\pi y/a)}\right] dxdy \qquad (12)$$

$$y = x'-x$$

We divide the integrand into two parts. First consider the asymptotic behaviour of,

$$\lim_{y\to -i\infty} \frac{e^{iy(k+n_0 2\pi/a-\pi/a)}}{\sin(\pi y/a)} = -\frac{i}{2} \lim_{y\to -i\infty} e^{iy(k+n_0 2\pi/a-2\pi/a)} \qquad (13)$$

According to our definition of $n_0$,

$$(k+n_0 2\pi/a - 2\pi/a) < 0 \qquad (14)$$

and (13) clearly vanishes in the limit. The other component of the integrand is more troublesome,

$$f'(y+x)e^{-ikf(y+x)} \qquad (15)$$

but we note that this is just the complex conjugate of the compressed wave. Despite being compressed it consists only of forward travelling waves. This does not guarantee that all the components wave vectors are positive: if negative frequencies are excited, a forward travelling have will have a negative wave vector. In the absence of negative frequencies all the components wave vectors, $k + ng$, are positive and their contributions to the contour vanishes as $y \to -i\infty$. In this case we can close the $y$ contour in the lower half plane,

$$N_{ph} = \frac{D_0^2}{\hbar\varepsilon_0 c_0} \frac{1}{k^2 a} \int_0^a ikf'(x)(-i)adx = \frac{aD_0^2}{\varepsilon_0 c_0 \hbar k} \qquad (16)$$

which we recognise as the number of photons captured in length $a$ of the incident wave, calculated in (4).

We stress that the theorem is violated if we excite negative frequencies which will have significance in the quantum context discussed below.

There is a second theorem which is always obeyed. If we count photons with negative frequency as having negative energy we obtain a different series,

$$\tilde{N}_{ph} = \sum_{n=-\infty}^{+\infty} \frac{a|d_n|^2}{\hbar\varepsilon_0 c_0 (k+n2\pi/a)}$$
$$= \frac{D_0^2}{\hbar\varepsilon_0 c_0} \sum_n \frac{(k+n2\pi/a)}{k^2 a} \iint e^{ik(f(x)-f(x'))-i(k+n2\pi/a)(x-x')} dxdx' \qquad (17)$$

which can also be summed to give,

$$\sum_{n=-\infty}^{+\infty} (k+n2\pi/a)e^{-i(k+n2\pi/a)x} \underset{\lim N\to\infty}{=} i\frac{d}{dx} \sum_{n=-N}^{+N} e^{-ikx-in2\pi x/a}$$
$$\underset{\lim N\to\infty}{=} i\frac{d}{dx} e^{-ikx} \frac{\sin((2N+1)\pi x/a)}{\sin(\pi x/a)}, \quad 0 < k < 2\pi/a \qquad (18)$$

Now we can always close the contour and in the limit we have,

$$\tilde{N}_{ph} \underset{\lim N\to\infty}{=} \frac{D_0^2}{\hbar k a \varepsilon_0 c_0} \iint e^{ik(f(x)-f(x'))} i\frac{d}{dx} e^{-ik(x-x')} \frac{\sin((2N+1)\pi(x-x')/a)}{\sin(\pi(x-x')/a)} dxdx'$$
$$= \frac{D_0^2}{\hbar k \varepsilon_0 c_0} \int_0^a f'(x)dx = \frac{a D_0^2}{\varepsilon_0 c_0 \hbar k} \qquad (19)$$

and $\tilde{N}_{ph}$ is conserved unconditionally. In the next section we demonstrate the validity of the theorems by direct computation.

We argue that when a negative frequency is excited the discrepancy between the two theorems shows that two photons must be added to the system: we take away the negative energy one and add it back with positive energy so the discrepancy between the unphysical zero sum theorem and the violated physical sum over positive energy is two photons. Of course this must be the case if we are to conserve momentum.

### 4. Illustrative calculations of photon number distribution over frequencies

We return to a specific model of a time dependent system as extensively investigated in previous papers by ourselves and others. It consists of a simple generalisation of a Bragg grating of the form shown in Eq.(1), moving with velocity $c_g = \Omega/g$. We stress that material comprising the grating does not move, rather the local properties are modulated in the synchronised form given above. This allows the structure to move synthetically with any velocity, unrestricted by the speed of light. This model has been widely adopted in time dependent studies of 'space-time crystals' [9] and of non-reciprocal systems [10,11]. Closely related models have been used to study topological aspects of so called time-crystals [12]. Our transfer matrix simulations use the following parameters,

$$\alpha_\varepsilon = \alpha_\mu = 0.05, \quad g = \Omega = 0.07,$$
$$\varepsilon_0 = \mu_0 = c_0 = 1.0 \qquad (20)$$

so that the grating velocity, $c_g = \Omega/g = c_0 = 1.0$, lies in the centre of the trans-luminal region. We choose $\omega = 10.1 \times \Omega$. The transfer matrix calculations presented in Fig. 3 are for transmission through two different thicknesses of grating corresponding to 32 (A,C) and 64 (B,D) spatial periods respectively, and show the photon content of each Fourier component of the transmitted wave alongside the energy content. As thickness increases photons are spread over more frequencies and when the spread reaches into negative frequencies the first theorem is violated, but in the two instances shown here by a very small amount: 0.07% in the case of the greater thickness. The second theorem is obeyed to machine precision as it must be.

The oscillations in Fourier component amplitude seen in Fig. 3 can be explained by reference to Fig. 1. Compression of the phase is determined by $f'$ which squeezes the original uniform phase oscillations, creating many Fourier components. Loosely speaking we can associate each point in the compressed space with a particular Fourier component according to how much the original wave is compressed at that point. Certainly it is a vicinity of this point which makes the most contribution. It will also be seen from Fig. 1 that in each period there are two points of equal compression each contributing to the same Fourier component but with a difference relative phase according to which part of the period they were harvested from. The points of equal compression start at the boundaries of the period and converge to the centre where $f'(x)$ is a maximum and the relative phase zero. The number of oscillations is therefore determined by the phase change over one period which is $ka/2\pi = k/g = 10.1$. Hence we see 10 oscillations.

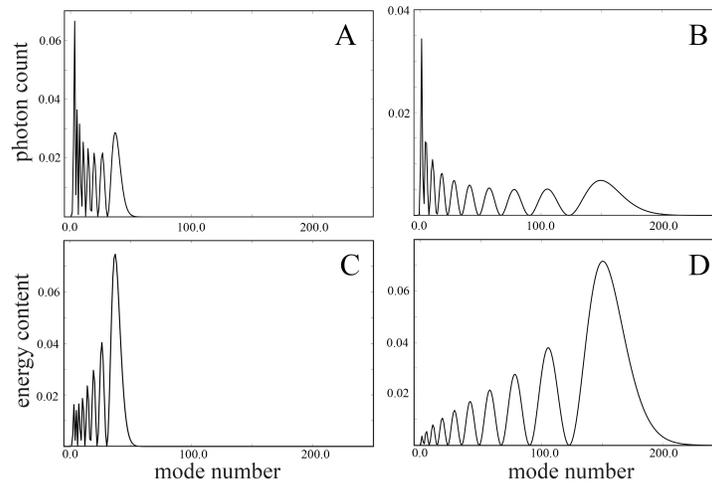

Fig. 3. Mode occupation numbers after a plane wave passes through a grating comprising **A** 32 periods of the grating and **B** 64 periods. **C** and **D** show the relative energy content for 32 and 64 periods respectively.

Finally we check the impact of breaking strict impedance matching. In Fig. 4 we show calculations for the same system as for Fig. 3 except that only the permittivity is modulated. Fig. 4A is to be compared to Fig. 3B and shows less diffusion of the photons to higher order modes, but still obeys the conservation law to within a fraction of a percent. This is evidenced by Fig. 4B which shows the number of photons in each backscattered mode. Violation of photon conservation is associated with backscattering which is zero for the impedance matched case but still very small at 0.014% in this more general scenario.

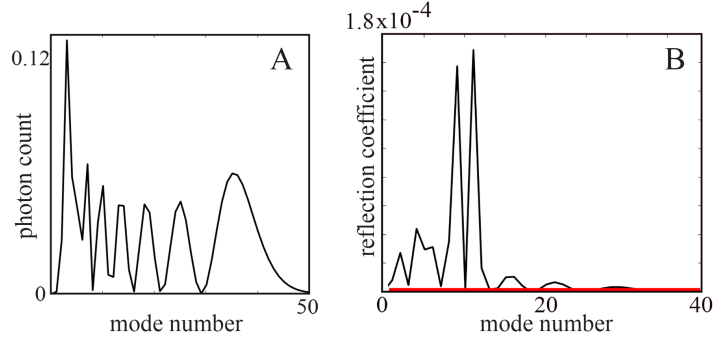

Fig. 4. Occupation numbers plotted against mode number after a plane wave passes through a grating comprising 64 periods. Fig. 4A is to be compared to Fig. 3B calculated for the same parameters except that for Fig. 4A the permeability is kept constant. Fig. 4B shows occupation of the backscattered modes which is very small indicating negligible violation of photon conservation. For the impedance matched case (red online) back scattering is zero to machine accuracy.

To summarise: $\tilde{N}_{ph}$, which counts negative frequencies as having negative energies, is obeyed to machine precision provided that the impedance matched, no reflection, condition is met. The true photon number, $N_{ph}$, is conserved only if no negative frequencies are present but in the examples provided here, is violated by very small amounts.

## 5. Photon conservation from a QED perspective

Casting the problem in terms of quantum electrodynamics addresses photon conservation in its own language, so to speak, and shows very directly when and how it is achieved.

We consider a 1D version of Maxwell's equations with waves propagating along the $x$ axis. Polarisation is preserved by the permittivity and permeability which are functions only of $x,t$ and periodic in $x$ with period $a$. Therefore we need consider only one polarisation obeying,

$$\frac{\partial E}{\partial x} = \frac{\partial B}{\partial t} = \frac{\partial}{\partial t}\left(Z^{-1}\mu\right)(H'), \quad \frac{\partial H'}{\partial x} = Z\frac{\partial D}{\partial t} = \frac{\partial}{\partial t}(Z\varepsilon)E, \quad (21)$$

$$H' = ZH, \quad \mu' = Z^{-1}\mu, \quad \varepsilon' = Z\varepsilon$$

where $H$ is oriented along the $y$ axis, $E$ along the $z$ axis, and $Z$ is a constant which we shall define later. We work with the vector potential in the Coulomb gauge,

$$E(x,t) = -\dot{A}(x,t), \quad H'(x,t) = -\partial_z A(x,t) \quad (22)$$

and $A$ obeys,

$$\frac{\partial}{\partial x}\frac{\partial_x A}{\mu'(x,t)} = \frac{\partial}{\partial t}\varepsilon'(x,t)\dot{A} \quad (23)$$

Lagrangian is,

$$L = \int \frac{1}{2}\left[\varepsilon'(x,t)\dot{A}^2 - \frac{(\partial_x A)^2}{\mu'(x,t)}\right]dx \quad (24)$$

$\pi(x,t)$ is the conjugate momentum obtained from the Lagrangian density,

$$\pi(x,t) = \frac{\partial L}{\partial \dot{A}} = +\varepsilon'(x,t)\dot{A}(x,t) \tag{25}$$

and on substituting into the Lagrangian,

$$L = \int \frac{1}{2}\left[\frac{\pi^2(x,t)}{\varepsilon'(x,t)} - \frac{(\partial_x A)^2}{\mu'(x,t)}\right]dx \tag{26}$$

We expand,

$$\partial_z A(z,t) = \sum_n \left[\begin{array}{l} \text{sgn}(\omega+n\Omega)(k+ng) \\ \times\left\{\begin{array}{l} A_{cn}\cos\left[\text{sgn}(\omega+n\Omega)(k+ng)z - |\omega+n\Omega|t\right] \\ +A_{sn}\sin\left[\text{sgn}(\omega+n\Omega)(k+ng)z - |\omega+n\Omega|t\right]\end{array}\right\}\end{array}\right],$$

$$\pi(z,t) = \sum_n \left\{\begin{array}{l}+\pi_{cn}\cos\left[\text{sgn}(\omega+n\Omega)(k+ng)z - |\omega+n\Omega|t\right] \\ +\pi_{sn}\sin\left[\text{sgn}(\omega+n\Omega)(k+ng)z - |\omega+n\Omega|t\right]\end{array}\right\} \tag{27}$$

where we use cos and sin to ensure that all quantities remain real. The expression $\text{sgn}(\omega+n\Omega)$ addresses the difficulty in QED that negative frequencies must be interpreted as positive energies. This we address by inverting the sign of both the frequency and the wave vector thus retaining the reality of a wave headed in the same direction as before the inversion and the benefit that we can continue to interpret $\hbar\,\text{sgn}(\omega+n\Omega)(k+ng)$ as the momentum and $\hbar|\omega+n\Omega|$ as the energy of a photon. The Lagrangian follows,

$$L = \frac{1}{2}\sum_{bnb'n'}\left[\left(\varepsilon'^{-1}\right)_{bnb'n'}\pi_{bn}\pi_{b'n'} - \text{sgn}(\omega_n\omega_{n'})k_n k_{n'} A_{bn}A_{b'n'}\left(\mu'^{-1}\right)_{bnb'n'}\right] \tag{28}$$

which we use to construct the Hamiltonian,

$$H = \sum_n \left[\pi_n \dot{A}_n\right] - L$$
$$= \frac{1}{2}\sum_{bnb'n'}\left[\left(\varepsilon'^{-1}\right)_{bnb'n'}\pi_{bn}\pi_{b'n'} + \text{sgn}(\omega_n\omega_{n'})k_n k_{n'} A_{bn}A_{b'n'}\left(\mu'^{-1}\right)_{bnb'n'}\right] \tag{29}$$

The subscript '$b$' refers to '$\sin$' or '$\cos$'. Next the conjugate variables are associated with operators as follows,

$$\frac{1}{(2\hbar\omega_{bn})^{1/2}}[\pi_{bn} + k_{bn}A_{bn}] \to \hat{a}_{bn},$$

$$\frac{1}{(2\hbar\omega_{bn})^{1/2}}[\pi_{bn} - k_{bn}A_{bn}] \to \hat{a}_{bn}^\dagger,$$

$$\pi_n \to \left(\frac{\hbar\omega_{bn}}{2}\right)^{1/2}\left[\hat{a}_{bn} + \hat{a}_{bn}^\dagger\right], \tag{30}$$

$$k_{bn}A_{bn} \to \left(\frac{\hbar\omega_{bn}}{2}\right)^{1/2}\left[\hat{a}_{bn} - \hat{a}_{bn}^\dagger\right]$$

Substituting into the Hamiltonian,

$$\hat{H} = \frac{\hbar}{4} \sum_{bb'} \sum_{nn'>0} (\omega_{bn}\omega_{b'n'})^{1/2} \begin{bmatrix} +\left[\left(\varepsilon'^{-1}\right)_{bnb'n'} + \left(\mu'^{-1}\right)_{bnb'n'}\right]\left[\hat{a}_{bn}\hat{a}^\dagger_{bn'} + \hat{a}^\dagger_{bn}\hat{a}_{bn'}\right] \\ +\left[\left(\varepsilon'^{-1}\right)_{bnb'n'} - \left(\mu'^{-1}\right)_{bnb'n'}\right]\left[\hat{a}_{bn}\hat{a}_{bn'} + \hat{a}^\dagger_{bn}\hat{a}^\dagger_{bn'}\right] \end{bmatrix}$$
$$+ \frac{\hbar}{4} \sum_{bb'} \sum_{nn'<0} (\omega_{bn}\omega_{b'n'})^{1/2} \begin{bmatrix} +\left[\left(\varepsilon'^{-1}\right)_{bnb'n'} - \left(\mu'^{-1}\right)_{bnb'n'}\right]\left[\hat{a}_{bn}\hat{a}^\dagger_{bn'} + \hat{a}^\dagger_{bn}\hat{a}_{bn'}\right] \\ +\left[\left(\varepsilon'^{-1}\right)_{bnb'n'} + \left(\mu'^{-1}\right)_{bnb'n'}\right]\left[\hat{a}_{bn}\hat{a}_{bn'} + \hat{a}^\dagger_{bn}\hat{a}^\dagger_{bn'}\right] \end{bmatrix} \quad (31)$$

This Hamiltonian is consistent with our conclusion from the classical equations concerning systems for which,

$$Z\varepsilon = \varepsilon' = \mu' = Z^{-1}\mu \quad (32)$$

Since $Z$ is an arbitrary constant any system in which $\varepsilon$ is everywhere proportional to $\mu$ satisfies this requirement and when the condition $nn' > 0$ is met as in the first summation in (31), creation and annihilation are always paired and photon conservation holds. Photon creation can still occur through transitions from positive to negative frequencies. In the quantum case this implies that a synthetically moving structure would spontaneously emit radiation even when impedance matched. The significance of impedance matching relates to black hole radiation where in the vicinity of the Schwarzschild singularity the effective values of permittivity and permeability are impedance matched and therefore according to our theorems radiation can only occur if negative frequencies are included. Of course, if a system is not impedance matched then there is no strict theorem. However the theorem will hold approximately provided that back scattering is minimal.

We note the resemblance of this positive to negative transition to the case of quantum friction where photons are generated when a Doppler shift moves frequencies across a positive/negative boundary [23,24].

**6.  Spontaneous emission of radiation**

The system can add photons to a field already present, but does this include the presence of vacuum fluctuations? This is a question closely related to emission of Hawking radiation [25] and to the several model systems proposed for mimicking the effect that may possibly be realised on a laboratory scale [26-32]. It is generally believed that Hawking radiation from black holes will never be observed so the only hope for experimental confirmation lies in model systems.

A connection with the Schwarzschild metric for a black hole can be established by following the path outlined in our earlier papers and making a Galilean transformation to a co-moving frame in which the grating is stationary. We refer the reader to these earlier papers for derivation of the constitutive relations in the new frame [12,14],

$$\varepsilon_{mov} = \frac{\varepsilon(X)}{1-\varepsilon(X)\mu(X)c_g^2}$$

$$\mu_{mov} = \frac{\mu(X)}{1-\varepsilon(X)\mu(X)c_g^2} \quad (33)$$

$$\xi_{mov} = -\frac{\varepsilon(X)\mu(X)c_g}{1-\varepsilon(X)\mu(X)c_g^2}$$

where $X = x - c_g t$. Closer examination of the co-moving parameters reveals pathological behaviour for grating velocities in the vicinity of the speed of light. The constitutive parameters in the co-moving frame show a singularity if there exists a point $X_s$ such that,

$$1 - \varepsilon(X_s)\mu(X_s)c_g^2 = 0 \quad (34)$$

Here the local velocity of light relative to the grating is zero and a pulse of light would continuously slow down on approach and never pass that point.

The condition for this transluminal region to occur in the impedance matched case is that,

$$1/\sqrt{1+2\alpha} < \sqrt{\varepsilon_1}\, c_g/c_0 < 1/\sqrt{1-2\alpha} \quad (35)$$

Within this range of grating velocities PT symmetry breaks down and energy can be extracted from the synthetic motion of the grating.

This is highly reminiscent of the effective values of permittivity and permeability calculated near the Schwarzschild radius. It has been pointed out that light propagating under the Schwarzschild metric [31] behaves as if in a medium with,

$$\varepsilon = \mu = \left(1+\frac{r_s}{r}\right)^3\left(1-\frac{r_s}{r}\right)^{-1}, \quad r_s = \frac{2GM}{c_0^2} \quad (36)$$

where $G$ is the gravitational constant and $M$ is the mass of the black hole. The singularity in the spatial component of the metric is of the same order as that in our transformed system.

The Galilean frame has a double singularity dividing space into two parts: in one part $c_\ell > c_g$ and in the other $c_\ell < c_g$. Light travelling in the forward direction cannot pass between the two but piles up intensity at the singularity where $c_\ell$ and $c_g$ converge. However the bianisotropic nature of the Galilean frame ensures that backwards travelling light sees no singularity and passes freely.

Experimental efforts are proceeding apace, but are likely to modulate only the permittivity. This helps the cause of generating Hawking radiation because net creation of photons from the ground state is allowed.

Further discussion of Hawking radiation is deferred to a subsequent paper.

### 7. Acknowledgements

E.G. acknowledges funding from a Junior Fellowship of the Simons Society of Fellows (855344). P.A.H. acknowledges funding from Fundação para a Ciencia e a Tecnologia and Instituto de Telecomunicações under project UIDB/50008/2020 and the CEEC Individual program from Fundação para a Ciencia e a Tecnologia with reference CEECIND/02947/2020. J.B.P. acknowledges funding from the Gordon and Betty Moore Foundation

**Disclosures**

The authors declare no conflicts of interest.

**References**

1. E. Galiffi, R. Tirole, S. Yin, H. Li, S. Vezzoli, P.A. Huidobro, ... & J.B. Pendry, "Photonics of time-varying media", arXiv preprint arXiv:2111.08640 (2021)
2. E.S. Cassedy, & A.A. Oliner, "Dispersion relations in time-space periodic media: Part I—Stable interactions", Proceedings of the IEEE, 51(10), 1342-1359 (1963).
3. F. Biancalana, A. Amann, A.V. Uskov, & E.P. O'Reilly, "Dynamics of light propagation in spatiotemporal dielectric structures", Physical Review E, 75(4), 046607 (2007w).
4. R. Fleury, D.L. Sounas, C.F. Sieck, M.R. Haberman, & A. Alù, "Sound isolation and giant linear nonreciprocity in a compact acoustic circulator", Science, **343**, 516-519 (2014).
5. Z.Yu, & S. Fan, "Complete optical isolation created by indirect interband photonic transitions", Nature photonics, **3**, 91-94 (2009).
6. S. Taravati, N. Chamanara, & C. Caloz, "Nonreciprocal electromagnetic scattering from a periodically space-time modulated slab and application to a quasisonic isolator", Physical Review B **96**, 165144 (2017).
7. D.L. Sounas, & A. Alu, "Non-reciprocal photonics based on time modulation", Nature Photonics, **11**, 774-783 (2017).
8. E. Galiffi, P.A. Huidobro, & J.B. Pendry, "Broadband nonreciprocal amplification in luminal metamaterials", Physical Review Letters, **123**, 206101 (2019).
9. X. Wen, X. Zhu, A. Fan, W.Y Tam, J. Zhu, H. Wu, ... & J. Li, "Unidirectional amplification with acoustic non-Hermitian space− time varying metamaterial", Communications Physics, **5**, 1-7 (2022).
10. J.B. Pendry, E. Galiffi, & P.A. Huidobro, "Gain mechanism in time-dependent media", Optica, **8**, 636-637 (2021).
11. M. Cromb, G.M. Gibson, E. Toninelli, M.J. Padgett, E.M. Wright, & D. Faccio, "Amplification of waves from a rotating body", Nature Physics, **16**, 1069-1073 (2020).
12. P.A. Huidobro, E. Galiffi, S. Guenneau, R.V. Craster, & J.B. Pendry, "Fresnel drag in space–time-modulated metamaterials", Proceedings of the National Academy of Sciences, **116**, 24943-24948 (2019).
13. E Galiffi, P.A. Huidobro, & J.B. Pendry, "An Archimedes' Screw for Light", arXiv preprint arXiv:2109.14460 (2021).
14. P.A. Huidobro, M.G. Silveirinha, E. Galiffi, & J.B. Pendry, "Homogenization theory of space-time metamaterials", Physical Review Applied, **16**, 014044 (2021).
15. X. Xu, Q. Wu, H. Chen, H. Nassar, Y. Chen, A. Norris, ... & G. Huang, "Physical observation of a robust acoustic pumping in waveguides with dynamic boundary", Physical Review Letters, **125**, 253901 (2020).
16. R. Fleury, A.B. Khanikaev, & A. Alu, "Floquet topological insulators for sound", Nature Communications, **7**, 1-11 (2016).
17. A. Darabi, X. Ni, M. Leamy, & A. Alù, "Reconfigurable Floquet elastodynamic topological insulator based on synthetic angular momentum bias", Science advances, **6**, eaba8656 (2020).
18. A. Dutt, M. Minkov, I.A. Williamson, & S. Fan, "Higher-order topological insulators in synthetic dimensions", Light: Science & Applications, **9**, 1-9 (2020).
19. E. Lustig, Y. Sharabi, and M. Segev, "Topological aspects of photonic time crystals" Optica, **5** 1390-5 (2018).
20. C. Caloz, & Z.L. Deck-Léger, "Spacetime metamaterials—part I: general concepts", IEEE Transactions on Antennas and Propagation, **68**, 1569-1582 (2019).
21. J.B. Pendry, E. Galiffi, & P.A. Huidobro, "Gain in time-dependent media—a new mechanism" JOSA B, **38**, 3360-3366 (2021).
22. E. Galiffi, M.G. Silveirinha, P.A. Huidobro, & J.B. Pendry, "Photon localization and Bloch symmetry breaking in luminal gratings" Physical Review B, **104**, 014302 (2021).
23. J.B. Pendry, "Shearing the Vacuum - Quantum Friction", J. Phys. [Condensed Matter], **9**, 10301-20 (1997).
24. J.B. Pendry, "Quantum Friction-Fact or Fiction?", New Journal of Physics, **12**, 033028 (2010 ).
25. S.W. Hawking, "Particle Creation by Black Holes", Commun. Math. Phys., **43**, 199-220 (1975).
26. W.G. Unruh, "Experimental Black-Hole Evaporation?" Physical Review Letters, **46**, 1351-53 (1981).
27. P.D. Nation, J.R. Johansson, M.P. Blencowe, & F. Nori, "Colloquium: Stimulating uncertainty: Amplifying the quantum vacuum with superconducting circuits", Reviews of Modern Physics, **84**, 1 (2012).
28. Scott J. Robertson, "The Theory of Hawking Radiation in Laboratory Analogues", Journal of Physics B: Atomic, Molecular and Optical Physics, **45**, 162001 (2012).
29. J. Sloan, et al. "Casimir Light in Dispersive Nanophotonics", Physical Review Letters, **127**, 053603 (2021).
30. J. Sloan, N. Rivera, J.D. Joannopoulos, *et al.* "Controlling two-photon emission from superluminal and accelerating index perturbations:, Nat. Phys. **18,** 67–74 (2022).
31. F. deFelice, "On the Gravitational Field Acting as an Optical Medium", General Relativity and Gravitation, **2**, 347-357 (1971).

32. R. Schützhold, G. Plunien, and G. Soff, "Dielectric Black Hole Analogs", Phys. Rev. Lett., **88** 061101-1 (2002).